# Effective mass calculations of SrTiO$_3$-based superlattices for thermoelectric applications lead to new layer design

Wilfried Wunderlich[1)], Hiromichi Ohta[2)], Kunichi Koumoto[2)],

1) Tokai University, Fac.Eng., Material Science Dep., Kitakaname 1117, Hiratsuka-shi, 259-1292, Japan
2) Nagoya University, Dept. Molecular Design Eng., Furo-cho, Chikusa, Nagoya, 464-8603, Japan

The effective mass is one of the main factors for enlarging the Seebeck coefficient and electronic conductivity of SrTiO3-based thermoelectric materials [1,2]. The goal of this paper is to clarify, how superlattices can change the effective mass and other features of the bandstructure. The natural Ruddlesden-Popper phase (SrTiO$_3$)$_n$(SrO)$_m$ with n=2, m=1 the situation changes, because the TiO$_6$-octahedrons are slightly extended, due to diluted density of the SrO layer. Another effect is the deformed electron density, which leads to reduced effective mass perpendicular to the layer, but enlarged parallel to the plane [3]. The average value of the effective mass over this anisotropy of the 2-dimensional electron gas (2DEG) for pure Ruddlesden-Popper phases is smaller, but can increase beyond the value of pure Pervoskite for certain doping elements.

In the same way, artificial superlattices (SrTiO$_3$)$_x$/(SrTi$_{1-z}$(Nb)$_z$O$_3$)$_y$ were examined. When fine nanostructures (n=2, m=3 or n=3, m=2) are present the effective mass increases, when the structure becomes coarser (n=4, m=1) smaller values are determined. These ab-initio calculation results of the increased effective mass have a great impact on development of futher nano-structures, as it is expected that the total figure-of-merit increased, due to the channeling of electrons in that layers of the superlattice where the electron mobility is higher and the thermal conductivity reduced.

**Introduction**

Heavily doped n-type semiconductor SrTiO$_3$ has been successfully developed as thermoelectric materials with rather large figure of merit of 0.34 at 1000K [1]. The reason for the rather high Seebeck coefficient is explained by enhancement of the effective mass [1-3], when heavy elements are doped on the *B*-(Ti-)site of SrTiO$_3$-based perovskites or when the lattice constant increased. The present study extends these investigations to layered perovskites, which are record holders for almost any property of functional materials, e.g. high $T_c$-superconductor, microwave resonators, piezoelectrics as well as giant magnetoresistance (GMR) materials. The reason is the confinement of the two-dimensional electron gas (2DEG) caused by embedding the electric conducting perovskite-blocks in between non-conducting layers with shielding function. In similar way, it has been shown [2-6], that layered perovskites are candidates for advanced thermoelectric materials, because the insertion of a SrO layer in between SrTiO$_3$ reduces the thermal conductivity $\kappa$ and hence increases the power-factor $S^2\sigma/\kappa$..

SrO containing layered SrTiO$_3$ perovkites are known as Ruddlesden-Popper (RP)-phases [5-8], which are thermodynamically stable phases in the phase diagram between SrO and SrTiO$_3$ [7,8]. The two most important species in the homologous series with different SrTiO$_3$-fractions *n* in the compound (SrTiO$_3$)$_n$(SrO)$_m$ are (SrTiO$_3$)SrO and (SrTiO$_3$)$_2$(SrO), abbreviated as STO214 and STO327 and are stable at room temperature. While doping with Nb increases the effective mass of SrTiO$_3$ up to m*/m0=10 [1], the effective mass at Ruddlesden popper phase STO327 remains only m*/m$_0$=3 even when the Nb doping is increased [6]. The reason is the asymmetric expansion of the TiO$^{6-}$-octahedra, which are compressed to cubic shape in perovskite, towards their original elongated shape with lower energy as in anatase or rutile. The SrO has a less packed density and allows this distortion. In particular, as the a and b-lattice constants of STO214 and STO327 are only little smaller than SrTiO$_3$ (0.3902 instead of 0.3905nm) and the atomic position at two joint SrTiO$_3$ layers in STO327 are the same as in the single crystal, but at the SrO/SrTiO$_3$ interface atomic relaxations take place, which are suppressed in the highly symmetric perovskite. In detail, the formerly in-plane Ti- and O-atom shift 0.011 and 0.055nm towards SrO, and the outermost O of the $TiO_6$-octaedron even 0.0360nm in the same direction, while the Sr atoms are pulled 0.151nm towards the $SrTiO_3$-layer, resulting in a quite wavy SrO-layer and elongated $TiO_6$-octaedron, which is almost as distorted as in anatase or rutile. These relaxations are almost equivalent for STO214 and STO327 and it is obvious, that these atomic relaxations will be present also in larger cells (n>3) or artificial cells (m>1), as studied in this study. After understanding this phenomenon, crystal engineering could restore the large effective mass by suitable doping of elements with large atomic radii [5]. Before further possibilities for engineering are explored e.g. in double-perovskites [9,10], layered structures need to be understood and this paper has such an intention.

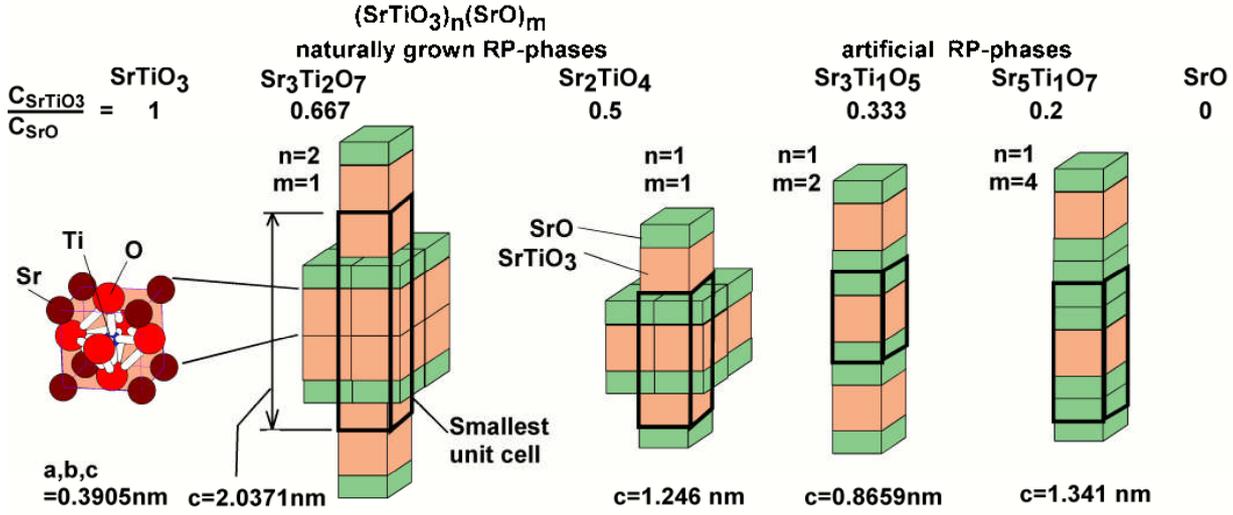

Fig. 1. Sequence of layers in $(SrTiO_3)_n(SrO)_m$-superlattices, the amount of $SrTiO_3$ is decreasing from the left to right as marked, starting from pure $SrTiO_3$, then the natural Ruddlesden-Popper phases with m=1, artifical superlattices with n=1, and finally SrO. The thick layer corresponds to $SrTiO_3$, the thin layer SrO in (110) orientation.

Recent MBE and PLD technology can also produce artificially RP-phases or superlattices with doped and undoped $SrTiO_3$ [4]. While La-doping occupies the Sr-site and decreases the effective mass [1,2], Nb occupies the Ti-site and in both cases the electron concentration increases linearly with the doping level, when the oxygen concentration is kept constant [1,2]. An icrease in effective mass increases the Seebeck coefficient and hence the performance of any thermoelectric device [9]. The effective mass can further be enhanced, when the conditions of a two-dimensional electron gas (2DEG) are realized, or in order words, the deformations of the electronic bands at an interface become so strong, that the high potential barrier forces the electrons to proceed only along the interface plane. The transport properties of $(SrTiO_3)_x/(SrTi_{1-z}(Nb)_zO_3)_y$ with z=0.2 have been experimentally measured in detail [4]. The electric conductivity decreases with increasing x-values, but the Seebeck coefficient increases and saturates for x>14 when y is kept constant (y=20). On the other hand the Seebeck coefficient increases, when y is decreased and x is kept constant (x=17). This paper clarifies these findings on atomistic scale by ab-initio simulations and gives further suggestions for improving.

**Calculation method**

Supercells of peroskite layers were generated by adding several unit cells of $SrTiO_3$ using the usual crystallographic data of $SrTiO_3$ (*Pm-3m*, $a_0$=0.3905nm), as well as STO214 (*I 4 /mmm*, a= 0.3902nm, c= 1.246nm), STO327 (*I 4 /mmm*, a= 0.3902nm, c=2.0371nm) and SrO (*F m -3 m*, a=0.5139nm). As the space group is reduced in symmetry, the redundant path in reciprocal space for becomes quite complicated. In order to make the results on layered persovskites comparable with those on simple perovskites, we use for the path in *k*-space the same sequence of reciprocal points as in $SrTiO_3$, the equivalent notation for extended perovskites are explained in [2] in detail. Due to the different supercell dimensions in directions of each of the unit vectors, the path in reciprocal space in-plane and out-of-plane become different in length. As SrO has alternating layers of Sr and O, in natural Ruddlesden Popper phases the perovskite blocks are shifted one half of the unit cell in the each of the two in-plane directions. In the case of artificial superlattices $(SrTiO_3)_n(SrO)_m$ with m=2 it is obvious that the second SrO layer will restore this shift as shown in fig. 1, and upper and lower perovskite blocks are aligned to each other.

The ab-initio software program Vasp [12] was used for calculations, which treats the multi-particle problem of electrons in a period crystal by local density approximation (LDA) according to the density-functional theory (DFT) using the pseudo-potential-method ($E_{cutoff}$=-380eV), respectively. For each simulation the band structure and density of states (DOS) were calculated. While for modeling of doping 2x2x2 symmetric supercells were used [2,3], for modeling superlattices supercells with other sizes (5x1x1 and 3x2x1) were used. When comparing calculations of a single unit cell and supercells of undoped $SrTiO_3$ the band-curvature and hence the effective mass is unchanged; the only difference is that some of the degeneracy is suppressed, especially when the supercell shape becomes tetragonal or orthorhombic (e.g. 5x1x1 or 3x2x1). Nevertheless, retrieving the main issues of the bandstructure even in unconventional supercells is considered as a successful benchmark test of the applied Vasp-software [12]. As for any crystals containing transition elements calculated by DFT-LDA the band-gap shows slightly smaller values than in experiments, which is usually adjusted by an appropriate value of the self-energy $U$ using the LDA+$U$ method. In this study there is no need to use that because the curvatures of bands are not affected and comparative calculations were performed.

The effective band masses $m_{Bi}$ were derived from the curvature of each band $i$ and extreme point $j$ in k-space (usual Γ point), namely the conduction band minimum or valence band maximum by fitting this a parabolic graph to the band [11]

$$\frac{1}{m_{B,i,j}} = \frac{1}{\hbar} \cdot \frac{d^2 E_{i,j}}{dk_{i,j}^2} \quad (1).$$

Their averages over $j$, namely heavy (h) and light (l) electron masses at different k-space points from the same band $i$ were calculated from

$$m_{B,i} = \left( m_{B,i,h}^{3/2} \cdot m_{B,i,l}^{3/2} \right)^{2/3} \quad (2).$$

As described in detail elsewhere [3], the mass of the electron closest in energy to the band-gap determines the electric properties. Hence, the effective mass $m^*$, which is relevant for macroscopic properties like the electric conductivity due to the Drude equation, was estimated from

$$m^* = m_e^* = m_{B,i} \quad (3)$$

with $i=$ 1 minimal conduction band for electrons. In the case of doped $SrTi_{1-x}Nb_xO_3$ for x>0.2 the calculated Fermi-levels lies about 250meV above the conduction band minimum. As the doping concentrations in the experiment are slightly lower, we can use the curvature of the band with lowest energy for calculation of the effective mass. Furthermore, due to Nb-doping the electron concentration is increased and in experiments this means a shift of the chemical potential, or in other words a reduction, at which oxygen vacancies are introduced. Both effects leads to an increase of the effective mass of up to 10 in experiments [1] which is also verified in the calculations [3]. In the case of superlattices the effective electron mass is different for in-plane and out-of-plane direction. The data of $m^*/m_0$ obtained from the bandstructures in fig. 2 for un-doped $SrTiO_3$ and the related superlattices according to fig. 1 are summarized as averages due to (2) in table 1. When such polycrystalline RP-material is measured in experiments, the arithmetic average over the two in-plane- and the out-of-plane-directions is considered as the corresponding effective mass measured in conduction experiments and in such a way the polycrystalline data in table 1 were calculated. The calculations for pure $SrTiO_3$ shows a value of $m_{B,h}/m_0$ =4.2 for heavy electrons around the Γ-point in Γ−Δ (100) direction, while the masses in Γ−Σ (110) or Γ−Λ (111) directions are smaller ($m_{B,l}/m_0$ =1.2), yielding to $m^*/m_0$= 4.8 in good agreement to [3] (table 1).

**Results and Discussion**
Effective masses of $SrTiO_3$- based Ruddlesden Popper and related lattices

The density-of-states of $SrTiO_3$ shows large density-of-states near the band gap in both, the valence as well as conduction band due to oxygen 2p and Ti-eg bands, respectively (fig. 2). In the superlattices, when the SrO amount is increased, the bandgap becomes larger and the DOS of the conduction band decreases. The bandstructure as shown in fig. 3 in the out-of-plane direction only has large values of effective masses around the Γ-point in Γ−Δ (100) direction, but the masses in Γ−Σ (110) or Γ−Λ (111) directions are smaller than one ($m_{B,l}/m_0$ =0.5), so that effective mass in out-of-plane direction becomes 2.94 smaller than in $SrTiO_3$, as shown in table 1. For STO214 the situation is similar, but it has a larger effective mass around the Γ-point in Γ−Δ (100) direction, so that the out-of-plane value 5.9 is larger than in $SrTiO_3$. direction in the out-of-plane direction is achieved, as shown in table 1. For the artificial superlattices STO315 the mass in out-of-plane increases further (7.0), as can be seem in fig. 3c as parallel band over the entire region Γ−Δ-X direction in reciprocal space. The reason for this enlargement is due to the two layers of SrO in between the $SrTiO_3$ block, which obviously have a large shielding effect and leading to large potential walls for the two-dimensional electron gas. When the amount of SrO is increased further, the electron density in the conduction band in diluted further due to the lack of Ti-2g orbitals. This leads to less degeneracy, the bands are more separated as can be seen on fig. 3 d and the stronger curvature leads to a decrease of the effective mass 3.5 in out-of-plane direction (table 1). It should be noted that all these values of effective mass represent the un-doped case. The average over the effective mass predicts values for polycrystalline material, which are in all cases smaller than the value of $SrTiO_3$. As mention in the introduction, the reason for this is the constriction of the $TiO^{6-}$-tetrahedron in the perovskite lattice into a symmetric shape, while they usually are elongated along one axis as in rutile or anatase.

Tab. 1. Values of effective mass $m^*/m_0$ of undoped $(SrTiO_3)_n(SrO)_m$-superlattices, as calculated by (1)-(4) from the bandstructure simulations in fig. 3.

|         | in-plane | out-of-plane | Average |
|---------|----------|--------------|---------|
| Sto     | 4.20     | 4.20         | 4.20    |
| Sto327  | 0.26     | 6.0          | 1.44    |
| Sto214  | 0.28     | 10.0         | 3.12    |
| Sto315  | 0.34     | 20.0         | 6.81    |
| Sto517  | 0.36     | 12.0         | 3.80    |
| SrO     | 0.51     | 0.51         | 0.51    |

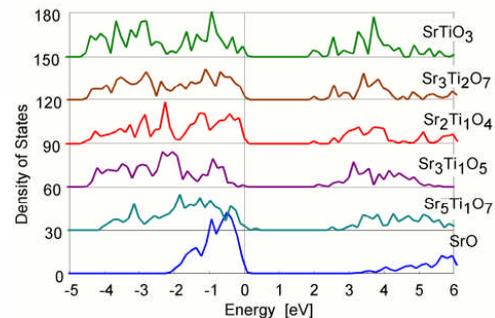

Fig. 2. Density of states n $(SrTiO_3)_n(SrO)_m$-superlattices (Ruddlesden-Popper phases) in the corresponding sequence as in fig. 1. The bandgap increases with the decreasing amount of $SrTiO_3$.

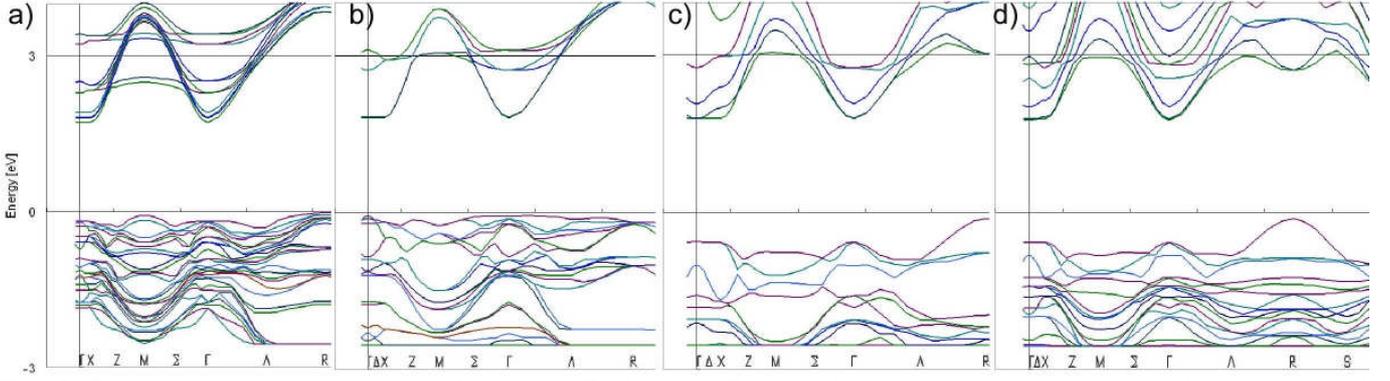

Fig. 3. Bandstructure calculations for the natural Ruddlesden-Popper phases shown in fig. 1 a) and b) and artificial superlattice as shown in fig 1 c) and d).

Concerning the desired increase of effective mass, it can be stated, that the two insulting layers of SrO have the largest effect. For Nb-doping the effective mass values increase due to increase of electron density in the conduction band and the oxygen vacancies. This has been confirmed for the doped Nb-$SrTiO_3$ lattice [2].

Nb-doped $SrTiO_3$-layers and strings embedded in $SrTiO_3$

Artificial superlattices of alternating $SrTiO_3$ layers with and without Nb-doping are promising candidates for enhanced properties due to the confined electron gas. Supercells in the geometry as shown in fig. 4 were created with dimensions a-f) 1x1x5 and g-l) 2x1x3. Two un-doped cases (fig. 4a and 4g) were calculated for comparison, in all other cases the Ti atoms in perovskite were partly substituted by Nb in different pattern as shown in fig. 4. The overall concentration of Ti-substitution by Nb increases from the left to the right as shown. To make the results comparable, we use the same notation as in [4], namely $(SrTiO_3)_x(SrTi_{1-z}Nb_zO_3)_y$ superlattices, where y is the thickness of the conducting doped $SrTi_{1-z}Nb_zO_3$ monolayers layers, marked with t in fig. 4b, and x the thickness of pure $SrTiO_3$, or in other words the distance of the conducting layers, marked with d in fig. 4b). The doping ratio z was 0.2 in [3], here it is z=1, but the case in fig. 4h) can be considered as a representation of z=0.5 with t=1ML and d=2ML on atomic scale. Similarly the case in fig. 4i) can be considered as z=0.5 with t=2ML and d=1ML. The case in fig. 4j) can be both, either a variation of case 4i) (z=0.5 with t=2ML and d=1ML) or a case z=1 with t=1ML one conducting layer with a distance of d=2ML. Due to the applied periodic boundary conditions in all three dimensions, the case shown in fig, 4b) leads to t=1ML thick $SrNbO_3$-layers with a distance of 4 ML embedded in $SrTiO_3$, the case in 4c) to 2ML thick Nb-layers with a distance of 3 ML, the case in fig. 4d) to 3ML thick Nb-layers with a distance of 2ML, the cases in fig.4e, f) to doped layers with varying space. Case 4h) in the 2x1x3 supercells can be also interpreted as one quantum string of $SrNbO_3$ embedded in $SrTiO_3$. The genuine structures with alternating $SrNbO_3$ and undoped $SrTiO_3$ layers are the cases in fig. 4b-d), the case in fig. 4j) with t= 1ML $SrNbO_3$ and d= 2ML and the case in fig. 4k) with t=1ML, d =ML.

The electronic bandstructure is shown in fig. 5 in the same sequence corresponding to fig. 4. In the un-doped cases fig. 5a) and g) the Fermi energy lies slightly above the valence band and the conduction band is unoccupied. In all Nb- containing cases the Fermi levels lies about 250meV above the minimum of the conduction band almost independent on the Nb-concentration, but the effective mass changes with the Nb-amount as shown in table 2. The out-of-plane direction as shown for each case in the lower bandstructure, shows only for Nb-concentrations less than 0.2 (fig. 5h) a large mass, in other cases a band with strong curvature made a "break-through" with a lower energy. This is in agreement with the critical doping value of x=0.2 in $SrTi_{1-x}Nb_xO_3$ found for best thermoelectric performance [1] and the highest effective mass $m^*/m_0$ >10 in [3] before the metal-to-insulator transition. Furthermore, as a general rule it can be stated, that all ordered cases (4b, c, d, j, l), where the Nb-atoms are aligned, large effective masses in the out-of-plane direction are observed. In the other cases, when the Nb-atoms are in more or less in random order, the effective mass is smaller. The cases of mono-atomic Nb layers t=1 with small spacing d=1 or 2 (fig. 5 j, l) have the highest effective masses.

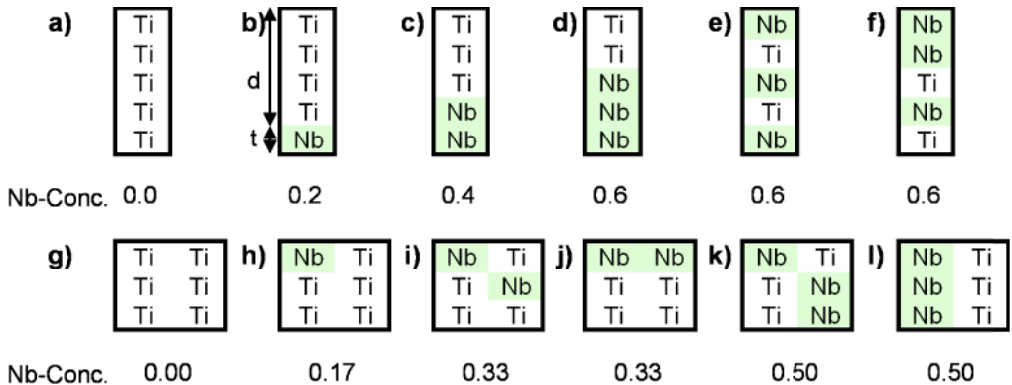

Fig. 4. Supercells with dimension 1x1x5 (upper row) and 2x1x3 (lower row) for calculation of $(SrTiO_3)_x(SrTi_{1-z}Nb_zO3)_y$ superlattices with z=1 with different thicknesses t of conducting layers and different distances.

Tab. 2. Values of effective mass $m^*/m_0$ of of $(SrTiO_3)_x(SrTi_{1-z}Nb_zO_3)_y$ superlattices as calculated from the bandstructure simulations in fig. 5. The corresponding structures are sketched in fig. 4.

| case | a) | b) | c) | d) | e) | f) |
|---|---|---|---|---|---|---|
| in-plane | 4.8 | 5.3 | 9 | 8 | 6 | 6 |
| out-of-plane | 4.8 | 0.2 | 0.18 | 0.3 | 0.3 | 0.3 |

| case | g) | h) | i) | j) | k) | l) |
|---|---|---|---|---|---|---|
| in-plane | 4.8 | 5 | 5 | 10 | 4 | 20 |
| out-of-plane | 4.8 | 5 | 0.2 | 0.2 | 0.2 | 0.2 |

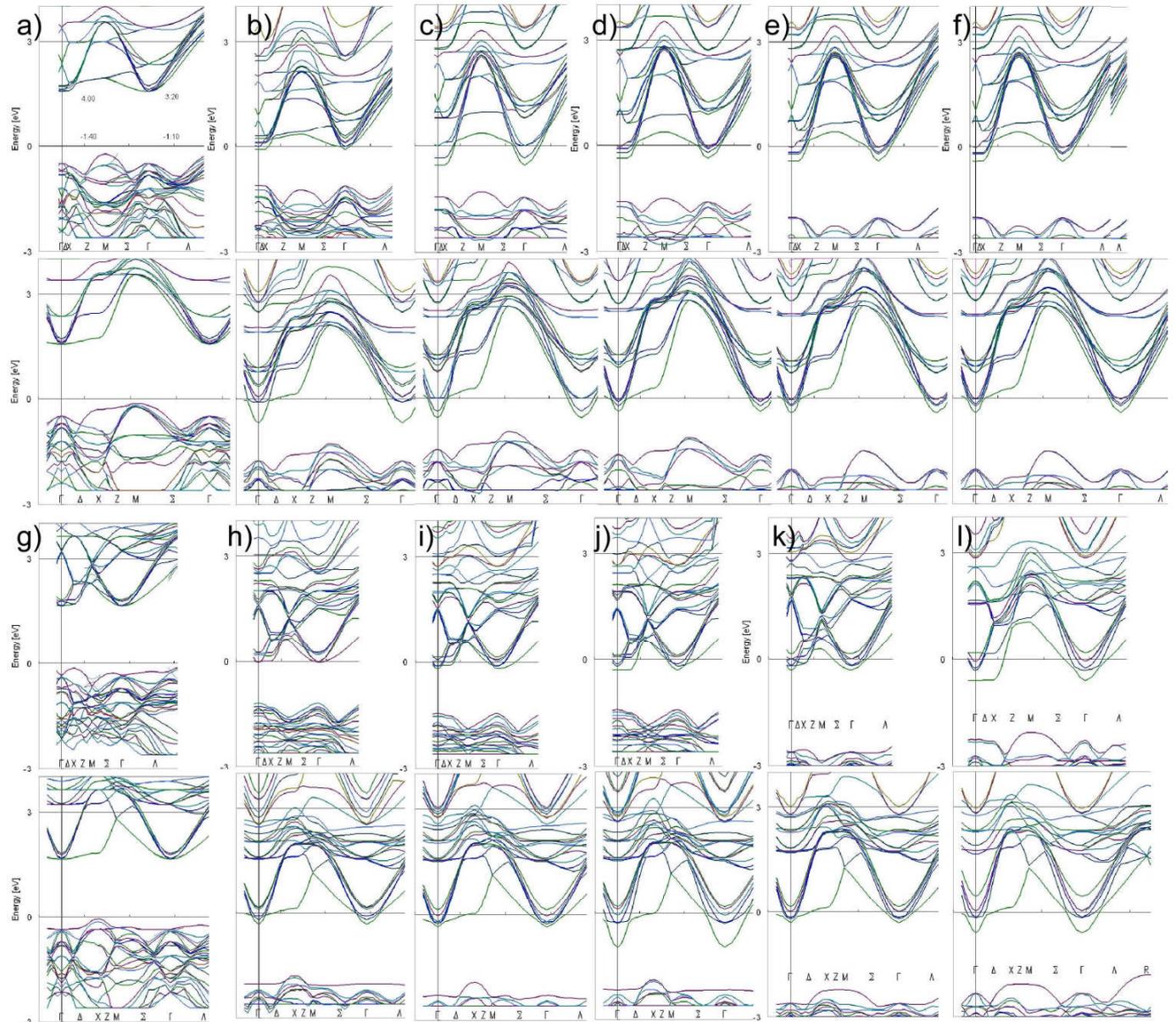

Fig. 5 Electronic bandstructure for the supercells shown in fig 4 in the same order. For each case the upper row shows the direction out-of-plane, the lower row the in-plane direction.

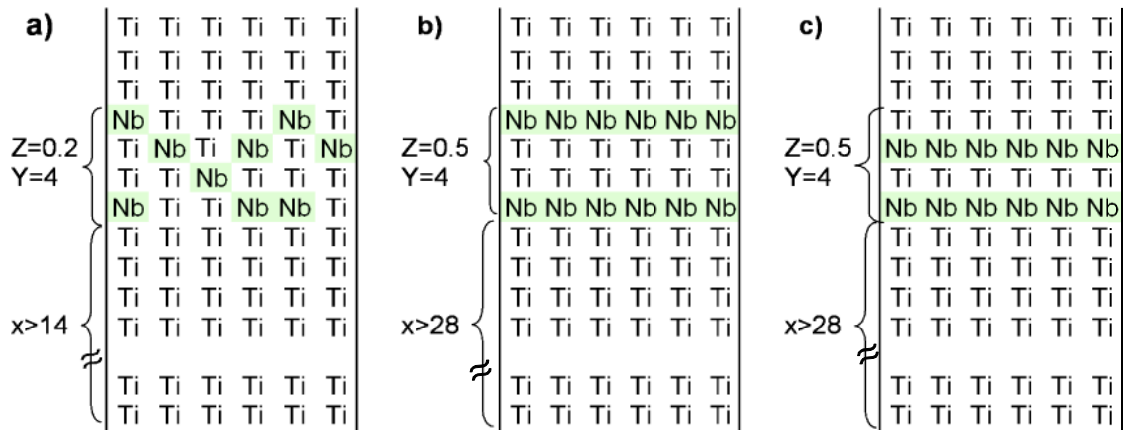

Fig. 6 Design of Nb-doped SrTiO$_3$-superlattices, a) as realized in [4], b), c) as suggested by this paper.

Summarizing all calculations in this paper, it can be stated, that mono-atomic conductive layers separated with 2ML thick insulating layers showed for both superlattices (SrTiO$_3$)$_n$(SrO)$_m$ and (SrTiO$_3$)$_x$/(SrTi$_{1-z}$(Nb)$_z$O$_3$)$_y$ the highest effective masses. The corresponding experimental results in [4] unsed for the conducting layers in (SrTiO$_3$)$_x$/(SrTi$_{1-z}$(Nb)$_z$O$_3$)$_y$ the presently best known bulk doping concentration of z=0.2. The calculations presented in this paper show however one additional aspect. Instead of random arrangement of the Nb and Ti atoms in the conduction layer to acchieve the doping ratio z=0.2 (fig. 6a), an ordering would even increase the effective mass, as can be seen in the cases in fig. 6b or c. According to the present simulation results, the best choice seems to be a ratio of one or two conducting layers of SrNbO$_3$ and two insulating layers of SrTiO$_3$ (case in fig. 4d) as a set. Two atomic layers correspond also well to the simulations of the SrO –SrTiO$_3$ superlattices in fig. 3c and the effective distance of electrons in any nano-scale experiment. The design of artificial superlattice can hence be optimized accoding to fig. 6 c or d, but it should be noticed that the overall concentration near the optimum value of SrTi$_{0.8}$Nb$_{0.2}$O$_3$ seems also to be important. To maintain this value, the number of insulating layers presumably need to be increased from y>14 to y>28. In other words, the prediction of this paper is, that the superlattice for the best thermoelectric performance with an optimum ratio of (SrTiO$_3$)$_x$/(SrTi$_{1-z}$(Nb)$_z$O$_3$)$_y$ with z=0.2 x= 1, y>14 [4] could be improved by (SrTiO$_3$)$_x$/((SrTiO$_3$)$_2$(SrNbO$_3$)$_3$)$_y$. with z=0.5 x=2, d=2, y>28. As the effective mass is the most important factor for increasing the Seebeck coefficient, further improvement also for other functional materials can be expected from such superlattice calculations.

**Summary**

Electronic bandstructure calculations using DFT-LDA were performed for different geometries of nano-layer structures of Nb-doped SrTiO$_3$ and Ruddlesden-Popper typer (SrTiO$_3$)$_n$(SrO)$_m$ superlattices. These calculations give important guidelines for manufacturing artificial superlattices and could clarify the following issues:
- While in Nb-doped SrTi$_{1-x}$Nb$_x$O$_3$ bulk specimens beyond the concentration of the metal-to-insulator transition at x=0.25, the effective masses drops from m$^*$/m$_0$ >8 to m$^*$/m$_0$ =0.2, in the case of SrTiO$_3$ with Nb-doped monolayers large effective masses can be maintained even for large Nb- concentrations (x=0.6), in the out-of-plane direction.
- The largest effective masses m$^*$/m$_0$= 20 were obtained for the cases, where conducting layers are thicknesses of t=1 or 2 ML, and distances of about d=2ML. The large effective mass is only in the out-of-plane direction, the in plane effective mass is as small as m$^*$/m$_0$=0.2. This results gives important suggestions for improving the design of nano-layered superlattices for thermoelectric applications.
- Even extended supercells could retrieve the main issues of the bandstructure, which is considered as a benchmark test of the applied Vasp-software.
- The effective mass derived from the bandstructure [3] corresponds well to experimental observations of effective masses in electric conductivity and Seebeck measurements [1].

**Acknowledgement**
The financial upport for this resreach project was provided from the Japanese Science and Technology agency in the framework of the JST-CREST project, which ic gratefully acknowledged.

**References**
[1] S. Ohta, T. Nomura, H. Ohta, and K. Koumoto, , J. Appl. Phys. 97 034106 (2005)
[2] Wilfried Wunderlich, Kunihito Koumoto, International Journal of Materials Research 97 (2006) 5 657-662
[3] Wilfried Wunderlich, Hiromichi Ohta, Kunihito Koumoto, cond-mat/0510013 (to be published)
[4] Kyu Hyoung Lee, Yoriko Muna, Hiromichi Ohta, and Kunihito Koumoto, Applied Physics Express 1 (2008) 015007 [5] Yifeng Wang, Kyu Hyoung Lee, Hideki Hyuga, Hideki Kita, Katsuhiko Inaba, Hiromichi Ohta and Kunihito Koumoto, Appl. Phys. Lett. 91 (2007) 242102
[6] Kyu Hyoung Lee Sung Wng Kim, Hiromichi Ohta and Kunihito Koumoto, J. Appl Phys 101 (2007) 083707
[7] Ruddlesden, S.N.; Popper, P. *Acta Crys.* 11 (1958) 54-55

[8] J.H. Haeni, C.D.Theis, D.G. Schlom, et.al. Appl. Phys. Lett. 78 [21] (2001) 3292
[9] Tatsuo Shimizu, Takeshi Yamaguchi, Appl. Phys. Lett. 85, 1167 (2004)
[10] Sugata Ray, Priya Mahadevan, Ashwani Kumar, D. D. Sarma, R. Cimino, M. Pedio, L. Ferrari, and A. Pesci Physical Review B 67 085109 (2003)
[11] A. Bulusu, D.G. Walker, Superlattices and Microstructures, Volume 44, Issue 1, July 2008, Pages 1-36
[12] G. Kresse, and J. Hafner, Phys. Rev. B 49 14251 (1994).